\documentclass[aps,preprint,prd,showpacs,nofootinbib]{revtex4}
\usepackage{amsmath}
\usepackage{graphicx}
\usepackage{dcolumn}
\usepackage{bm}
\usepackage{amssymb}
\usepackage{latexsym}

\def\be{\begin{equation}}
\def\ee{\end{equation}}
\def\ba{\begin{eqnarray}}
\def\ea{\end{eqnarray}}

\begin{document}

\title{Preheating and Dark Sector of Universe}

\author{Yun-Song Piao$^{a,b}$}
\email{yspiao@itp.ac.cn} \affiliation{${}^a$Institute of High
Energy Physics, Chinese Academy of Science, P.O. Box 918-4,
Beijing 100039, P. R. China} \affiliation{${}^b$Interdisciplinary
Center of Theoretical Studies, Chinese Academy of Sciences, P.O.
Box 2735, Beijing 100080, China}

\begin{abstract}

Regarding long life particles produced during preheating after
inflation as dark matter, we find that its back reaction on the
field $\varphi$ could lock $\varphi$ in a false vacuum up to
today. This false vacuum can drive the accelerated expansion of
universe at late time and play the role of dark energy.
When the number density of dark matter particles is dilute to some
value, the field $\varphi$ becomes tachyonic and rolls to its true
minima rapidly, and the acceleration of universe ceases. We
discuss the constraints on the parameters of model from the
observations of dark energy and dark matter halos on subgalactic
scale.

\end{abstract}

\maketitle

Recent data from observations have given a possible picture where
we live in a spatially flat universe consisting of predominantly
unknown dark sector. In particular, at present about 1/3 of the
energy density of universe behaves as pressureless matter, which
is not luminous and has litter interactions with the usual baryon
matter, dubbed dark matter. The remaining about 2/3 is named dark
energy, which drives the current accelerated expansion of
universe. Recently, relevant issues of dark matter and dark energy
have received increased attentions, see \cite{PR, P, C2, S, T} for
reviews.

The simplest form of dark energy is a small positive cosmological
constant, which can fit data nicely and might be
phenomenologically the most appealing choice. But in this case,
its value of cosmological constant has to be fine tuned extremely
to an incredible level, in the meantime the coincidence problem
is also required to explain. Furthermore, a constant positive
vacuum energy will lead inevitably to eternal acceleration of
universe and the existence of future causal horizon, which may be
most undesirable in string theory, because it inhibits the
construction of S-matrix. The alternative is Quintessence
\cite{WRP, ZWS}, which is a rolling light scalar field with a
normal dynamical term and not interacts with other matters. It
could terminate accelerated expansion by reaching the minimum of
its potential or by gravitational back reaction \cite{LLZB}. It is
generally assumed that due to some as yet not understanding
mechanism the fundamental vacuum energy of universe is 0. The dark
energy observed is dominated by some fields that have not yet
relaxed to their vacuum state. The scalar field involving
Quintessence is required to have a very shallow potential, which
can be make its evolution overdamped by the expansion of universe
until recently. But it suffers from certain problems \cite{C, KL}.
For generical potentials the field is nearly massless $m_{Q}\sim
h_0\sim 10^{-33}$eV, where $h_0$ is present Hubble parameter. Such
a small mass may be inconsistent with radiation corrections.

In this note, we attempts to provides a single theoretical
framework for dark sector of universe. The other ones with this
aim can be found in Ref. \cite{ZPC, BLZ, BBS}. Regarding long life
particles produced during preheating after inflation as dark
matter, we find that its back reaction on the field $\varphi$ can
lock $\varphi$ in a false vacuum up to today. This false vacuum
can play the role of dark energy at late time. When the number
density of dark matter particles is dilute to some value, the
field $\varphi$ rolls to its true minima and dark energy
disappears. This model avoids some problems mentioned above.
Further, dark matter particles produced during the nonthermal
production \cite{LHZB} could be relativistic, and their comoving
free streaming scales could be as large as of the order 0.1 Mpc,
which lead to a severe suppression of the density fluctuations on
scales less than the free streaming scale. Thus under certain
conditions the discrepancies between the observations of dark
matter halos on the subgalactic scales and the predictions of the
standard cold dark matter scenario could also be resolved in this
model.

The reheating theory is one of the most important parts of
inflation cosmology, which has been developed extensively several
years ago \cite{KLS}. During preheating after inflation,
parameteric resonance \cite{KLS} or instant preheating \cite{FKL}
will lead to the production of many particles. The universe is
reheated after these particles decay. But if particles have a long
enough life, the back reaction produced by them can significantly
affect the motion of oscillating inflaton, for example, which may
reduced to a temporary symmetry restoration for a double well
potential \cite{KLS2, KKLT}. Trapping moduli at enhanced symmetry
points by quantum production of light particles has been studied
in Ref. \cite{KLLMMS}. Trapping of a scalar field which has a
potential can lead to a short period of accelerated expansion in
situations with steeper potentials than would otherwise allow this
has been shown. We assume, for our purpose, that other particles
except $\chi$ particles produced during preheating decays into
radiation and reheats the universe, thus their backreactions can
not affect the shape of $\varphi$ potential. But $\chi$ particles
still surviving until today will lead to a correction to the
effective potential of $\varphi$ and make $\varphi =0$ become a
local false vacuum when its backreaction effect is enough large.
We expect that the energy density of this false vacuum could be
responsible for the observed accelerated expansion while the
$\chi$ particles produced could be regarded as dark matter, which
locks the $\varphi$ field in the false vacuum equal to the
cosmological constant observed. Since the number density of $\chi$
particles decreases with the expansion of universe, at some time
when its back reaction is no more large than the techyonic mass of
$\varphi$, the field $\varphi$ will be released and roll down to
its true minimum rapidly. Thus the system reaches true vacuum and
accelerated expansion of universe ceases, see Fig.1 for an
illustration.

\begin{figure}[t]
\begin{center}
\includegraphics[width=8cm]{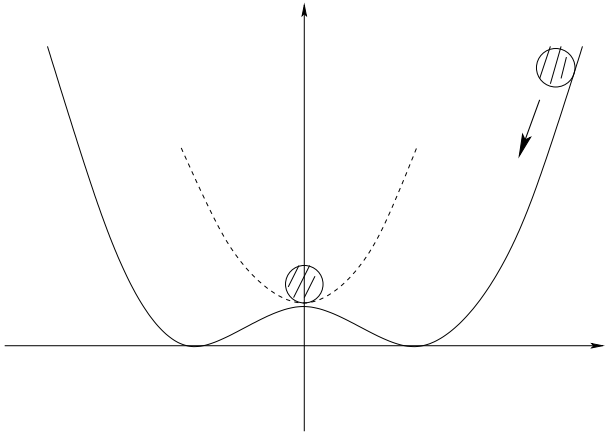}
\caption{ Illustration of model. Initially $\varphi >m_p$,
inflation occurs, and after inflation, the field will roll down
along its potential, see solid line. When the $\varphi$ field
reaches a small region about $\varphi =0$, the adiabaticity
condition will be violated and the productions of $\chi$ particles
leads to a correction to the potential of $\varphi$, see dashed
line, which makes $\varphi$ local in $\varphi =0$. The energy
density of false vacuum of $\varphi$ field and $\chi$ particles
produced are regarded as dark energy and dark matter respectively
in this model. }
\end{center}
\end{figure}

To implement this model, we consider such an effective potential
of field $\varphi$ as follows \be {\cal V}(\phi)= \left\lbrace
\begin{matrix}\alpha \varphi^4 &{\rm for}& \varphi\gg \varphi_* \cr
\beta(\varphi^2-\varphi_*^2)^2 &{\rm for}& \varphi \sim \varphi_*
\end{matrix} \right. \label{l}\ee
where the scalar field $\varphi$ has a chaotic inflation potential
for large $\varphi$, which is regarded as an asymptotic one here,
where inflation occurs, and has a double well potential for small
$\varphi$, whose global minima are at $\varphi = \varphi_*$, and
the $\varphi =0$ is an unstable saddle point, whose tachyonic mass
$\sim -\beta \varphi_*^2$. Suppose that an universe is initially
in inflation regime, where $\varphi \gtrsim m_p$, after inflation
ends, the field $\varphi$ will roll down and oscillate around
$\varphi = 0$ , then finally cease in some minimum of its
potential.

But when considering the interactions of $\varphi$ with other
scalar fields, the case will be different.
When the $\varphi$ field reaches a small region about $\varphi
=0$, the adiabaticity condition will be violated and the
productions of $\chi$ particles and other particles with fields
coupling to inflaton field $\varphi$ will occur. We consider an
interaction between a massive $\chi$ field ${1\over 2}m_{\chi}^2
\chi^2$ and inflaton as follows ${1\over 2}g^2\chi^2 \varphi^2$.
Thus the effective mass of $\chi$ field is $m_{\chi eff}^2=
m_{\chi}^2+g^2\varphi^2$, which decreases with the rolling down of
$\varphi$ after the end of inflation. When the field $\varphi$
arrives at about $\varphi =0$, where ${\dot{m}}_{\chi eff}\gtrsim
m_{\chi eff }^2$, relevant process becomes non adiabatic. For
$v\ll g$, $|\varphi| \lesssim (v/g)^{1/2}$ is in a narrow region,
where $v$ is the velocity of $\varphi$ about $\varphi =0$. Thus
the process of particle production occurs nearly instantaneously,
$\triangle t_*\sim (g v)^{-1/2}$. In this case the uncertainty
principle implies that the particles produced have typical momenta
$k\sim (\triangle t_*)^{-1} \sim (g v)^{1/2}$
Thus following \cite{KLS, FKL}, the occupation number $n_k$ of
$\chi$ particles with momenta $k$ suddenly acquires the value \be
n_k =\exp{(-{\pi(k^2+m_{\chi}^2)\over gv})} \label{nk}\ee This
value dose not change until the field $\varphi$ rolls to the point
$\varphi =0$ again. We expect that instant preheating and decay of
other particles may be very effective, which can make the universe
reheat to some temperature rapidly.
Though long life $\chi$ particles produced during initial
oscillation is negligible for the evolution of background, its
back reaction will provide a correction to the effective potential
of $\varphi$, which makes $\varphi =0$ become a temporal minimum.
During radiation dominated the $\varphi$ field may gently lands in
$\varphi =0$ without many oscillations with large velocity. Thus
in this case the parametric resonance may not occur and only a few
initial oscillations may be important. We simply estimate the
particle number density $n_{\chi}$ produced as \be n_{\chi}
={1\over 2\pi^2}\int dk k^2 n_k = {(gv)^{3/2}\over 8\pi^3}
\exp{(-{\pi m_{\chi}^2 \over gv})} \ee For $m_{\chi}^2 \ll gv$,
which is reasonable and can be seen from following calculations,
we have \be n_{\chi}\simeq {(gv)^{3/2}\over
8\pi^3}\label{nchi1}\ee

The kinetic energy of $\varphi$ field after reheating locked at
$\varphi =0$ is larger than its potential energy and is
relativistic. But since the $\chi$ particles has a bare mass, this
can make it become non relativistic before its energy overpasses
radiation. Considering the particle number density $n_{\chi}\sim
a^{-3} \sim T^{3}$, where we have neglected tiny difference
between coefficients during different periods, thus after the
$\chi$ particles become non relativistic, its energy density can
be written as \be \rho_{\chi}\sim n_{\chi}m_{\chi} ({T\over
T_r})^3 \label{rchi}\ee where $T_r$ is the reheating temperature
resulting from the decay of other particles produced during
preheating. From (\ref{l}) and (\ref{rchi}), that at present the
energy density of dark matter is approximately equal to dark
energy's requires \be n_{\chi}m_{\chi} ({T_0\over T_r})^3 \sim
\beta\varphi_{*}^4 \sim \Lambda \label{larl}\ee where $T_0$ is CMB
temperature and $\Lambda$ is the cosmological constant observed at
present. Furthermore, to make $\varphi$ stay in a false vacuum up
to today requires \be g^2\langle\chi^2\rangle \gtrsim \beta
\varphi_{*}^2 \label{larm}\ee For non relativistic $\chi$
particles, $\langle\chi^2\rangle$ can be reduced as \be
\langle\chi^2\rangle\simeq {1\over 2\pi^2} \int {n_k k^2 dk\over
\sqrt{k^2+m^2_{\chi}}} \simeq {n_{\chi}\over m_{\chi}}
\label{x2}\ee Thus substituting (\ref{x2}) into (\ref{larm}), one
obtain \be {g n_{\chi}\over m_{\chi}}({T_0\over T_r})^3 \gtrsim
\beta \varphi_{*}^2 \label{larm1}\ee

Combining (\ref{larl}) and (\ref{larm1}), \be m_{\chi}\lesssim g
\varphi_* \label{mchi}\ee is given. In this model, the energy
density of false vacuum is regarded as dark energy observed, thus
$\beta \varphi_*^4 \sim \Lambda \sim 10^{-120}m_p^4$, where $m_p$
is the Planck scale, which implies $\varphi_* \sim \beta^{-{1\over
4}} 10^{-30} m_p \label{varstar}$. Regarding the $\varphi$ field
as a modulus, a natural value of $\beta$ could be \be \beta\sim
({m_s\over m_p})^4 \label{beta}\ee where $m_s$ may be the
supersymmetry breaking scale. Thus one obtain $ \varphi_* \sim
10^{-30} m_p^2/ m_s $. Thus for $m_s\sim$Tev, $\varphi_*$ can be
Tev order, {\it i.e.} $\varphi_*\sim m_s$. In this case, one can
further obtain \be \Lambda \sim \beta \varphi_*^4 \sim ({m_s\over
m_p})^4 m_s^4 \label{mla}\ee which has been mentioned as a
possible suggestions for value of observed cosmological constant
\cite{AHKM}, where supersymmetry is assumed as breaking at Tev
scale by an order parameter chiral superfield which makes that
electroweak symmetry breaking become a direct consequence of
supersymmetry breaking. In their argument, the vacuum energy is
given by $(m_s^2/m_p)^4$.

Combining (\ref{larl}) and (\ref{mchi}), \be n_{\chi} \gtrsim
{\beta \varphi^3_*\over g} ({T_r\over T_0})^3 \label{nchi2}\ee is
given. Instituting it into (\ref{nchi1}), one obtain \be g\gtrsim
(8\pi)^{6\over 5} {\beta^{2\over 5} \varphi_*^{6\over 5}\over
v^{3\over 5}} ({T_r\over T_0})^{6\over 5} \label{gv}\ee During
initial oscillation after inflation the kinetic energy of
$\varphi$ at the bottom of valley is far larger than its potential
energy. Thus the velocity at $\varphi =0$ is hardly affected by
potential at small $\varphi$. For large $\varphi$, the potential
is $\alpha \varphi^4$. The numerical calculation shows $v\sim
0.01\sqrt{\alpha} m_p^2$ at $\varphi = 0$ for the first
oscillation. The proper fluctuation amplitude responsible for
large scale structure requires $\alpha \sim 10^{-13}$. Thus $v\sim
10^{-8} m_p^2 $. Considering (\ref{beta}), \be g\gtrsim 10^{-3}
({m_s\over m_p})^{14\over 5}({T_r\over T_0})^{6\over 5}
\label{g}\ee For example, taking $T_0\sim 10^{-13}$Gev and
$T_r\sim 10^{15}$Gev, we have $g\gtrsim 10^{-4}$.

The standard cold dark matter scenario predicts too much power on
small scale. Several possible resolutions have been proposed to
this apparent discrepancy \cite{LHZB, SS} It is known that below
the free streaming scale, the power spectrum can be severely
damped. The studies shows that to explain dark matter halos on the
subgalactic scales, the free streaming scale should be $\sim 0.1$
Mpc \cite{CAV}. The comoving free streaming scale $R_f$ for
nonthermal particles can be calculated as \cite{BMY, LHZB} \ba & &
R_f =\int_{t_i}^{t_{eq}} {u(t^\prime)\over a(t^\prime)}dt^\prime
\simeq \int_0^{t_{eq}} {u(t^\prime)\over a(t^\prime)}dt^\prime
\nonumber \\ &\simeq &  2r t_{eq} (1+ z_{eq})^2
\ln{(\sqrt{1+{1\over r^2(1+z_{eq})^2 }}+{1\over r(1+z_{eq})})}
\label{rf}\ea where $r\equiv ak/m$ is defined, which is a constant
during the evolution of universe, and the subscript eq denotes
when the energy density of radiation equals to that of matter. For
$R_f\sim 0.1$ Mpc, from (\ref{rf}), one obtain $r\sim 10^{-7}$,
which gives rise to a further constraint on parameter of the
model. Instituting \be r\simeq {T_0\over T_r} {(g v)^{1/2}\over
m_{\chi}} \sim 10^{-7} \ee into (\ref{g}), we have \be T_r
\lesssim ({\sqrt{v}\over 10^{-8} \varphi_* })^{5/8} ({m_p\over
m_s})^{7/8} T_0 \sim 10^{26} T_0\sim 10^{13} {\rm Gev}
\label{tr}\ee Thus for $T_r \gg 10^{13}$ Gev, $r\ll 10^{-7}$,
which means that the $\chi$ particle produced during preheating
serves as a good candidate for cold dark matter. However, a lower
reheating temperature will be more required to solve the
subgalactic scale problem of cold dark matter. For example, taking
$T_r\sim 10^7$ Gev and $g\sim 10^{-2}$, we obtain $r\sim 10^{-7}$.

The $\chi$ particles will still not decay until today requires
that its decay rate satisfies the conditions $\Gamma_{\chi} < h_0
\sim\sqrt{\Lambda}/m_p$. The coupling $g_{\chi}$ of $\chi$ with
its decay products generally lies in the range
$m_{\chi}/m_p\lesssim g_{\chi} \lesssim 1$, where the lower bound
is given by the gravitational decay of $\chi$ particles. Thus for
$\Gamma_{\chi} \sim g_{\chi}^2 m_{\chi}$, the bound $m_{\chi}
\lesssim 10^{-20}m_p $ is obtained, which can be consistent with
(\ref{mchi}).

When the energy of false vacuum dominate dark matter's at late
time, the universe will expand more fast. Thus the number density
of dark matter particles will be diluted more rapidly and its back
reaction on the effective potential of $\varphi$ field will be
weakened more fast than that during radiation or matter dominated
periods. From (\ref{g}), at some time in the future, when CMB
temperature decreases to \be T \sim 10^{-3} ({m_s\over
m_p})^{7\over 3}{T_r\over g^{5\over 6}} \label{t}\ee $\varphi= 0$
will become an unstable saddle point. From (\ref{mla}), the
tachyonic mass of $\varphi$ about $\varphi =0 $ is
$m_{\varphi}\sim (m_s/m_p)^2 m_s\sim 10^{15} h_0\gg h_0$. Thus the
$\varphi$ field will roll to its true minima rapidly, where its
potential energy is 0. In this case the accelerated expansion will
cease and the universe will be eventually dominated by massive
$\chi$ particles with mass square $m_{\chi eff}^2=
m^2_{\chi}+g\varphi_*^2$. From (\ref{mchi}), we have $m_{\chi
eff}\simeq g\varphi_*$. The efficient reheating requires $g\gg
10^{-4}$, thus the $\chi$ particle will eventually gravitationally
decay, whose products will dominate the universe in the final.

From (\ref{mchi}), (\ref{beta}) and (\ref{mla}), we see that the
parameters $\beta$, $\varphi_*$ and $m_\chi$ can be connected to
only two natural mass scale $m_s $ and $m_p$ with $\alpha\sim
10^{-13}$ for a successful inflation and $g$ for an enough dark
particles produced. Therefor, we can be placed in a false vacuum
leaded to by the back reaction of dark matter particles without
special fine tuning.

In conclusion, we proposed a single model of dark sector of
universe. The long life particles produced during preheating after
inflation are regarded as dark matter, whose back reaction on
$\varphi$ field locks $\varphi$ in a false vacuum, which drive the
accelerated expansion of universe at late time. When the number
density of dark matter particles is dilute to some value, the
field $\varphi$ rolls to its true minima and the acceleration of
universe ceases. This model not only retains basic predictions of
standard dark energy and cold dark matter scenario, but avoids
eternal acceleration.
Further, for a lower reheating temperature, the problem of cold
dark matter on subgalactic scale is also cured. Moreover, since
$m_{\varphi}\sim 10^{15}h_0 \sim 10^{9} h_{eq}$, compared with
quintessence model,
the correction from supergravity is negligible.
In addition, this model may be also applied to where the potential
has rich vacuum structures at small $\varphi$. Trapping effect
reduced by particles production may help to give a dynamical
selection mechanism for vacua \cite{KLLMMS}. In this process one
with a little mass and long life among kind of produced particles
may play a role of dark matter, which traps $\varphi$ in some
vacuum with observed dark energy.

{\bf Acknowledgments} The author would like to thank Xiao-Jun Bi,
Bo Feng, Xinmin Zhang for helpful discussions, and specially,
Mingzhe Li for valuable comments. This work is supported by K.C.
Wang Postdoc Foundation.

\end{document}